\documentclass[12pt]{article}
\usepackage{graphicx,epsfig}
\usepackage{cite}
\usepackage{amsmath}
\usepackage{amsfonts}
\usepackage{amssymb}
\usepackage{wrapfig}
\newcommand{\no}{\nonumber\\}
\newcommand{\be}{\begin{equation}}
\newcommand{\ee}{\end{equation}}
\newcommand{\ba}{\begin{eqnarray}}
\newcommand{\ea}{\end{eqnarray}}

\date{}
\begin{document}
\title{
Natural cancellation of powerlike divergences in the Standard Model by dark matter through gauged Higgs portal}
\author{S.S. Afonin$^{1}$, A.A. Andrianov$^{1,2}$,
 V.A. Andrianov$^{1}$ and D. Espriu$^{2}$}
\maketitle
\begin{center}
\vspace{-3mm}
\small{$^{1}${\it  Department of Theoretical Physics,\\ Saint-Petersburg State University, 199034 St. Petersburg, Russia}}\\
\small{$^{2}${\it Departament  de F\'isica Qu\`antica i Astrof\`isica and Institut de Ci\`encies del
Cosmos (ICCUB),\\
Universitat de Barcelona, Mart\'\i \ i Franqu\`es 1, 08028 Barcelona, Spain}}
\end{center}
\maketitle
\begin{abstract}
We discuss the connection between the vacuum energy problem and the number of dark matter sectors and its relation to the cancellation of quadratic
divergences both in the cosmological constant and simultaneously in the polarization tensor renormalizing the Higgs meson mass.
In the absence of global supersymmetry, the cancellations of quartic and quadratic divergences call for a dark Higgs scalar sector. We estimate the number of hidden(dark) gauged
Higgs field  sectors to be equal to five. The masses of dark bosons may be of the order of the Standard Model boson masses.
Then the  following picture of the Standard Model emerges: 6 leptons and 6 quarks are accompanied by 6 boson sectors, only one of the latter is realized
in the conventional Standard Model.

\end{abstract}

\section{Introduction}
A natural compensation of the net effect of  powerlike ultraviolet divergences in vacuum energy  of the Standard Model (SM) is crucial to
extend its applicability to high energies and to the large scale and early Universe. This was already a concern for W. Pauli \cite{pauli,giul}
in the mid-last-century, and later for Ya. Zeldovich \cite{zeld}. Compensation of quadratic ultraviolet divergences in the Higgs meson masses
was also addressed by M. Veltman \cite{velt}, but after measurements of the top-quark and Higgs boson masses the hypothetical compensation of
quadratic divergences turned out to be unfeasible.

On the other side a reliable identification of those divergences seems to be a subtle problem as it involves a way of regularization of
divergent Feynman integrals \cite{rgstab,oswu, pass,kamen2}. Nevertheless the distribution of kinematical and discrete degrees of freedom
unifies the accounting of UV divergences \cite{kamen2,acgks,kamen3} and makes it less sensitive to a specific regularization.

It is known that  supersymmetry (if it exists) could suppress divergences by compensating contributions of fermions and bosons \cite{kaul}. However,
there are some difficulties with identification of superpartners in the real world at scales of the Standard Model. So we ignore this possibility.

In the 90s some of us undertook \cite{roman1,roman2} the task of fine-tuning of quartic divergences in the vacuum energy and in the quadratic
divergences in the Higgs boson mass originating from the loop contributions of fermions and bosons by modifying the Veltman  condition \cite{velt}.
The proposed modification used different cutoff scales for fermion $\Lambda_F$ and boson $\Lambda_B > \Lambda_F $ matter sectors.
In 1994 the prediction of a top-quark mass at $175\pm 5 $ GeV, made in \cite{roman1,roman2}, was confirmed experimentally with a good precision,
but the Higgs boson mass required for the correct fine tuning happened to be larger than it was measured in 2012 \cite{pdg}.

New attempts to cure the hierarchy problem were associated \cite{zee,bhatt,davu,wilcz} with the dark sector
of matter fields interacting to the SM via sterile operators (see reviews \cite{arc1,arc2,leb}). Among others, there were models
for the reduction of the cosmological constant \cite{kamen2,kamen3}, several versions of dark boson portals \cite{karahan, jamal, pukhov}, and
towers of dark bosons \cite{athr,boos} were built.

The usage of different scales for effective high energy range of fermion and boson physics was avoided. Instead, the dark boson
matter d.o.f.'s  helped efficiently in reduction of quadratic divergences in the modified Veltman condition.

The definition of coupling constants in the compensation conditions depends on energy scale. In principle, a specific choice must be
specified by the renormalization group. More universal conditions could be derived by requirement of coupling constant reduction
in a Kubo-Sibolt-Zimmermann style~\cite{ksz}.

\section{Cancellation of the vacuum energy density}
Following this way we start by analying  the vacuum fluctuations of quantum fields which contribute to the energy-momentum tensor,
the ground state components of which behave as a cosmological constant.
We seek for sources of reduction of a radiatively induced cosmological constant in the framework of the standard field-theoretical treatment
of zero point modes encoded in vacuum energy.
A generally accepted  renormalization of the vacuum energy does not exist, while a straightforward cutoff imposed on the integration of the
visible d.o.f. in four-momentum space  gives huge values, to be compensated by dark matter d.o.f.'s.

Recall that summing up the zero-point energies of all the normal modes of a field component of mass $m$ up to a wave number cutoff $K\gg m$
yields the vacuum energy density for each particle d.o.f. \cite{zeld}. To the leading order in $\hbar$, i.e. by considering free theories
and neglecting interactions, the kinematical parts of the individual d.o.f.'s uniformly read
\ba
\langle\,\rho\,(m,K)\,\rangle &=&\frac12\,\int_0^K\frac{4\pi k^2dk}{ (2\pi)^3}\,
\sqrt{k^2+m^2}\no
&=& \frac{K^4}{16\pi^2}+\frac{K^2m^2}{16\pi^2}-\frac{m^4}{32\pi^2} \left[\,\ln\frac{K}{m}-\frac14+\ln2+O\left(\frac{m}{K}\right)^2\,\right]\ ,
\label{zeropoint}
\ea
and for different d.o.f.'s look universally determined  in various  regularization schemes.
The quartic divergence in $K$ here contributes to the cosmological constant, but the quadratic divergence will also be in the focus of our work.
The expression~\eqref{zeropoint} contains a logarithmic divergence too, but we will not discuss this type of contribution in the present work
as it deserves a separate study (see, e.g., the discussions in \cite{kamen2,acgks,kamen3,visser}).

If we trust general relativity up to the Planck scale $M_p$ we might take $K\simeq M_p = (8\pi G_{\cal N})^{-1/2}\,,$ which would give
\be
\langle\,\rho\,\rangle\approx 2^{-10}\,\pi^{-4}\,G_{\cal N}^{-2}=2\times 10^{71}\ {\rm GeV}^4\ .
\ee
The genuine vacuum energy will collect the number of the corresponding flavour, color, spin and singlet parts of particles (fields)
with different signs for fermions (minus) and bosons (plus).
Let us consider the observable world consisting of all Standard Model particles and a certain set of dark matter (SM sterile) particles.
Their numbers in the Lagrangian add up to $N_{F,v} + N_{F,d}$ of visible and dark fermions, respectively, and  $N_{B,v} + N_{B,d}$ of visible and dark
bosons, respectively. In the SM we know the number of visible fermions and bosons
\ba
N_{F,v}&=&4N_c\times N_f \times N_g + 4N_g + 2N_g = 72 +12 + 6 = 90;\no
N_{B,v} &=& 2(N_W +N_Z +N_\gamma) +N_H + 2(N_c^2 -1) +2N_G = 8+4+16+2 = 30, \label{defic}
\ea
where $N_c$,$N_f$ and $N_g$  are the numbers of colors, flavors (in one generation) and generations; $N_W=2 ,N_Z=1, N_\gamma=1$ and $N_H=4$  are the numbers of $W,Z$
 bosons, photons and Higgs bosons in the gauged doublet before condensation, $N_G$ is a number of gravitons (i.e. one).
This counting is done in the Landau gauge with transversal polarizations for massive gauge bosons.
We notice that in  previous papers \cite{roman1,roman2} the graviton polarizations were not taken into account and that will be important
for building our model in what follows.

From \eqref{defic} one finds the number of boson d.o.f.'s missing in order to have compensation of quartic divergences
 $$N_{F,v} - N_{B,v} = 60 = N_{B,d}
 $$
$$N_{F,d} = \widetilde N_{B,d}$$ for the minimal set of SM fermions that must be accompanied by extra boson d.o.f.'s in order to balance a
substantially reduced vacuum density made of \eqref{zeropoint}. In order to balance powerlike divergences in visible and pure dark sectors
one will eventually need more dark bosons  $\tilde N_{B,d}$ which are not involved in the Standard Model observations (see below).

Let us conjecture the content of families of scalars and vector mesons needed to reach the most efficient compensation of $N_{F,v} - N_{B,v} = 60$
unbalanced d.o.f.'s achievable with the help of dark matter particles. Evidently one could involve 60 real singlet scalars but
this is not welcome if we would like to make the dark matter physics similar to the visible sector. Clearly, the most compact choice includes
five dark sets with blocks of 12 bosons; that is three massive  bosons $\times$ three polarizations plus one massless gauge boson (dark photon)
$\times$ two polarizations, plus one massive Higgs scalar singlet. We emphasize that this nice picture emerges by virtue of taking into
account the graviton d.o.f.'s.

 \section{Quadratic divergences: compensation in vacuum energy}
 As it follows from \eqref{zeropoint} the balance of zero-point energies  would be achieved if the Pauli relation \cite{pauli,giul} holds
\be \sum_{H_0} m_{H_0}^2 +  3\sum_{V} m_V^2 - 4 \sum_F m_F^2 =0\,? \label{pauli}
\ee
where $H_0$ stands for neutral scalar Higgslike fields and $V, F$ is for massive vector bosons and Dirac fermions.
The vector boson masses are assumed to be generated by the Higgs mechanism. In the minimal version fermions are only visible and dominated
by the top quark mass, $m_F \rightarrow m_t$.
In the Standard Model it reads
\be
  m_{H_0}^2 +  3(2m_W^2 + m_Z^2)\approx 12  m_t^2\, ?\label{smpauli}
\ee
But it does not hold in the Standard model. Instead, inclusion of dark sectors changes the quadratic relation to
\ba
 && m_{H_0}^2 +  3(2m_W^2 + m_Z^2)- 12  m_t^2 \approx\no && 4\sum_{\widetilde F} m^2_{\widetilde F} -\sum_{\widetilde H} m^2_{\widetilde H}
  -3\sum_{\widetilde V} m_{\widetilde V}^2 \approx \\&& \big[- 12 (172)^2 + 125^2 + 3 (2 (80)^2 + 91^2)  )\big]\times \text{GeV}^2=- 525^2\, \text{GeV}^2 , \label{smpaulimod}
\ea
where $\widetilde F, \widetilde H,\widetilde V$ denote dark particles. This quantity  must be distributed among dark particles.
The minimal symmetric choice of bosons is given by a model with five symmetric gauged  Higgs  doublets and without dark fermions.

Let us suppose the dark boson masses and dark gauge fields couplings to be of the same order.  Then the five dark sectors  could be taken
symmetric in masses and the discrepancy in \eqref{smpaulimod} is covered, for instance, by bosons with masses  $m_{\widetilde H} \simeq 104,6$
GeV; $m_{\widetilde W} \simeq 67$ GeV; $m_{\widetilde Z} \simeq 76$ GeV. Many other possibilities are certainly possible.

\section{Quadratic divergences: Higgs boson masses}
\begin{figure}
[!htb]
\vspace{-2.0cm}
\begin{center}
\includegraphics
[scale=0.8]
{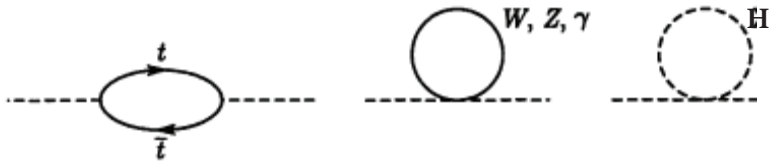}
\vspace{-19.0cm}
\caption{Quadratic divergences in radiative corrections for Higgs boson mass
}
\end{center}
\end{figure}
In the 90s some of us had studied \cite{roman1,roman2} the fine-tuning of quadratic divergences in the Higgs boson mass
originating from loop contributions of fermions and bosons (generalized Veltman condition):
\be
\Delta m^2_H = \frac{3}{16\pi^2}\, \left( \quad\int_{v}^{\Lambda_{F}}  \frac{d^4k}{k^2} 4g_t^2 -
\int_{v}^{\Lambda_{B}}\frac{d^4k}{k^2}\biggl\lbrace \frac34 g^2
+ \frac14 g'^2+
\lambda \biggr\rbrace \right)+ O\left(v^2 \ln(\Lambda^2/v^2)\right), \label{modveltman}
\ee
where common notations were used for the electroweak coupling constants $g,\,g'$, Higgs boson self-interaction  of fourth order
$\lambda$ and Yukawa coupling constant for the heaviest, $t$ quark, $g_t$. Eventually it is assumed all these constants
being effective, run with energy-momentum governed by the Renormalization Group.
By integration by parts, we can convince ourselves that the leading contribution is located in the vicinity of the upper limits
$\Lambda_F,\Lambda_B$. Let us retain the SM symmetries and the universality of gauge interactions for three fermion generations
as well as the SM symmetries for bosons up to the SM validity limits.  But we do not impose any supersymmetry requirements in the
UV region, i.e. we restrict our considerations to a non-supersymmetric SM particle set. Then the fermion and boson scales
seemed  \cite{roman1,roman2} to be inequivalent $\Lambda_F< \Lambda_B$ in order to  provide compensation of quartic and quadratic
UV divergences.

On the other hand, if SUSY would be present, an universal cutoff could be provided if one involves the dark matter sector for
reducing the vacuum energy imbalance.

Let us take into account the contribution of dark matter fields which are sterile with respect the SM interactions
but couple observable SM particles with dark matter. As we have a deficiency of bosons let us add just a number of SM sterile bosons
to equalize fermion and boson contribution in \eqref{modveltman}. One has several choices to saturate the modified Veltman relation:
a number of sterile Higgs-like bosons with or without sets of dark vector bosons. When keeping in mind the previous count of the required
sets of bosons to compensate quartic divergences in vacuum energy one concludes that possible choices might be combinations of a
number of dark Higgs doublets $\widetilde H_j$ supplemented with a subset of dark $SU(2) \times U(1)$ vector bosons; for instance,
$n, n\leq 5$ (gauged doublets + gauge field actions) + $15 - 3n$ ungauged Higgs complex doublets (the latter,of course, can be replaced
with a number of four Higgs real singlets).
\ba
&&4m^2_t=(2m^2_W+m^2_Z+m^2_H + \Delta m^2_{\widetilde H});
\nonumber\\
&& \Delta m^2_{\widetilde H}= \sum_{\widetilde{particle}} m^2_{\widetilde H} \no &&\simeq \big[4(172)^2 - (2 (80)^2 + 91^2  + 125^2)\big]\times \text{GeV}^2=286^2 \,\text{GeV}^2;
\label{extvelt}
\ea
at the RG scale $\sim 100$ GeV. This mass shift $\Delta m^2_{\widetilde H}$ must be distributed between masses of ungauged and gauged
dark Higgs doublets or singlets. In the minimal version it originates from the Higgs portal potential
$|H|^2 \sum_{\text{dark}} |\tilde H|^2$. We emphasize that it is not necessary to reveal all dark matter fields  through this portal in
the compensation mechanism of quadratic divergencies. As a result, quadratic divergences can be eliminated in the SM Higgs channel
but the associated dark matter scalars may suffer from large quadratic divergences themselves which enforce them to be rather supermassive
and not protecting the SM sector from  huge divergences. Evidently we have to supplement the complete lagrangian with a number of dark
fermions $N_{F,d}$ and additional dark bosons   $\widetilde N_{B,d} =  N_{F,d}$ so that the entire set of bosons  couple to one or more
dark fermions in order to provide the cancellation of quadratic divergences in all boson channels.

\section{Dark matter extension of SM}
The boson lagrangian of extended Standard model may take the following form:
\ba
{\cal L}_{\text{extSM}} &=& - \frac{1}{4g^2} W_{\mu\nu}^j W^{\mu\nu}_j -  \frac{1}{4(g')^2} B_{\mu\nu} B^{\mu\nu}\no &-& \sum_a\Big[\frac{1}{4g_a^2} \widetilde W_{\mu\nu\,a}^j \widetilde W^{\mu\nu,a}_j + \frac{1}{4(g'_a)^2}\widetilde B_{\mu\nu,a} \widetilde  B^{\mu\nu,a}\Big]\no &+& (\rm gauge \, fixings \, and \,Faddeev-Popov\, ghost\, determinants )
\no&&+ (D_\mu H)^\dagger D^\mu H - V(H^\dagger H)+\no
&+&\sum_a\Big[ (\widetilde D_\mu \widetilde H_a)^\dagger \widetilde D^\mu \widetilde H_a\Big]+\sum_b\Big[(\partial_\mu \widetilde h_b)^\dagger \partial^\mu\widetilde h_b\Big]+  V(H^\dagger H;\, \widetilde H^\dagger_a\widetilde H_a; \widetilde h^\dagger_b\widetilde h_b ),\\&&  V(H^\dagger H) = -m^2 H^\dagger H + \lambda (H^\dagger H)^2,\nonumber
\ea
where the vector triplets $\widetilde W_{\mu,a}, \widetilde B_{\mu,a}$, Higgs doublets $\widetilde H_a$ and Higgs singlets $\widetilde h_b$
stand for fields of a dark matter and the covariant derivatives $D_\mu \widetilde H_a$ contain the dark gauge fields. We notice that
gauge fixing and F.-P. ghost action eliminate unphysical components of vector fields and allow to count only physical polarizations in the
vacuum energy combination \eqref{defic} due to BRST invariance.



For the fermion sector of the SM, different choices of dark matter content are given by a composition of scalar potential,
\ba
&& V_\text{portal} = H^\dagger H \big[\sum_{a =1,\ldots,n} \widetilde\lambda_{0a}\widetilde H^\dagger_a\widetilde H_a + \sum_{b=1,\ldots,60 - 12n} \widetilde\lambda_{0b} \widetilde h_b\widetilde h_b \big];\no&&
V(H^\dagger H; \widetilde H^\dagger_a\widetilde H_a; \widetilde h_b\widetilde h_b ) = V_\text{portal} + V( \widetilde H^\dagger_a\widetilde H_a; \widetilde h_b\widetilde h_b ).
\ea
If the dark matter sector consists of gauged Higgs doublets $(n=5)$ such a model is maximally symmetric under permutation of Higgs doublet blocks including related dark gauge bosons. For this case let us assume the following pattern of dark Higgs potential:  equality of portal couplings $\widetilde\lambda_{0a} \equiv \bar\lambda, a = 1,\cdots,5$ and separability of dark matter clusters,
\be
V( \widetilde H^\dagger_a\widetilde H_a) = \sum_{a =1,\ldots,5} \big[-\widetilde\mu_a^2 \widetilde H_a^\dagger \widetilde H_a + \widetilde\lambda_a (\widetilde H_a^\dagger \widetilde H_a)^2 \big]
\ee
 A typical magnitude of  portal couplings between visible and dark sectors can be estimated to bring a positive $\Delta m^2_{\widetilde H}$. This is of course a leading-order contribution in $\hbar$ in the effective potential.

 \section{Summary}

We have proposed a model for the necessary content of visible and dark sector extending the SM in order to diminish the quartic divergences in vacuum energy.
For the visible fermion sector the maximal number of blocks of dark Higgs bosons accompanied by dark $SU(2)\times U(1)$ gauge bosons equals to five.
The generalized Veltman condition is enforced and it fits well when  supplemented by quadratic divergences through loops in the portal between visible and dark boson fields.
A class of the models with a special potential for dark Higgses is proposed that removes the quadratic divergences at least in the visible part of the SM.

The present work can be extended in various directions. First of all, the RG stability of the generalized Veltman condition should be investigated, its final
form is in progress.
Also more detailed predictions and bounds for the parameters of extended SM are under elaboration.


\section{Acknowledgments}
This work of S.S.A., A.A.A. and V.A.A. was supported by the Grant of Russian Science Foundation  21-12-00020.
D.E. acknowledges financial support from the State Agency for Research
of the Spanish Ministry of Science and Innovation through the “Unit of Excellence Maria
de Maeztu 2020-2023” award to the Institute of Cosmos Sciences (CEX2019-000918-M) and
from PID2019-105614GB-C21, 2017-SGR-929 and 2021-SGR-249 grants. We are grateful to O.O. Novikov  for fruitful discussions.

\end{document}